\documentclass[%
reprint,
showpacs,
amsmath,amssymb,
aps,prl,
]{revtex4-1}
\usepackage{braket}
\usepackage{float}
\usepackage{graphicx}% Include figure files
\usepackage{dcolumn}% Align table columns on decimal point
\usepackage{bm}% bold math
\usepackage{subfigure}
\usepackage{color}
\usepackage{ulem}

\begin{document}
\hyphenation{eq-ua-tions diff-er-ent only sce-nario also however equi-librium fila-ment results works re-mains two still account University} %  CNT CHARMM three single word list
% Use the \preprint command to place your local institutional report
% number in the upper righthand corner of the title page in preprint mode.
% Multiple \preprint commands are allowed.
% Use the 'preprintnumbers' class option to override journal defaults
% to display numbers if necessary
%\preprint{}

%Title of paper
\title {Transition to the ultimate regime in two-dimensional Rayleigh-B\'enard convection}

\author{Xiaojue Zhu$^{1}$}
\email{xiaojue.zhu@utwente.nl}
\author{Varghese Mathai$^{1}$}

\author{Richard J. A. M. Stevens$^{1}$}

\author{Roberto Verzicco$^{2,1}$}

\author{Detlef Lohse$^{1,3}$}
\email{d.lohse@utwente.nl}
\affiliation{$^1$Physics of Fluids Group and Max Planck Center for Complex Fluid Dynamics, MESA+ Institute and J. M. Burgers Centre for Fluid Dynamics, University of Twente, P.O. Box 217, 7500AE Enschede, The Netherlands\\
$^2$Dipartimento di Ingegneria Industriale, University of Rome `Tor Vergata',
Via del Politecnico 1, Roma 00133, Italy\\
$^3$Max Planck Institute for Dynamics and Self-Organization, 37077 G\"ottingen, Germany}

%\noaffiliation
%\homepage[]{Your web page}
%\thanks{}
%\altaffiliation{}
%\noaffiliation

%Collaboration name if desired (requires use of superscriptaddress
%option in \documentclass). \noaffiliation is required (may also be
%used with the \author command).
%\collaboration can be followed by \email, \homepage, \thanks as well.
%\collaboration{}
%\noaffiliation

%\date{\today}

\begin{abstract}

The possible transition to the so-called ultimate regime, wherein both the bulk and the boundary layers are turbulent, has been an outstanding issue in thermal convection, since the seminal work by Kraichnan [Phys. Fluids 5, 1374 (1962)]. Yet, when this transition takes place and how the local flow induces it is not fully understood. Here, by performing two-dimensional simulations of Rayleigh-B\'enard turbulence covering six decades in Rayleigh number Ra up to $10^{14}$ for Prandtl number Pr $=1$, for the first time in numerical simulations we find the transition to the ultimate regime, namely at $\textrm{Ra}^*=10^{13}$. We reveal how the emission of thermal plumes enhances the global heat transport, leading to a steeper increase of the Nusselt number than the classical Malkus scaling $\textrm{Nu} \sim \textrm{Ra}^{1/3}$ [Proc. R. Soc. London A 225, 196 (1954)]. Beyond the transition, the mean velocity profiles are logarithmic throughout, indicating turbulent boundary layers. In contrast, the temperature profiles are only locally logarithmic, namely within the regions where plumes are emitted, and where the local Nusselt number has an effective scaling $\textrm{Nu} \sim \textrm{Ra}^{0.38}$, corresponding to the effective scaling in the ultimate regime.

%We find that during the transition, the mean velocity profiles are indeed logarithmic, while for temperature, the logarithmic profiles can only be observed locally within the regions where plumes are emitted. In contrast to the conventional perception, these regions do not grow in their spatial extent, despite the increase in Ra. Therefore, the transition to the ultimate regime may not be explained as an increase in the fraction of plume-emitting regions of the boundary layer, but rather as the gradual takeover of the global heat transport by the contribution arising from these regions of plume ejection.
%It is the the gradual takeover of the heat flux originating from these plume ejecting regions, which triggers the transition.
%, characterized by an increase in the Nusselt vs. Rayleigh scaling exponent (Nu $\sim$ Ra$^{\beta}$).

%\begin{description}
%\end{description}
\end{abstract}

% insert suggested PACS numbers in braces on next line
%\pacs{47.20.Ky, 46.40.Jj, 47.63.Gd, 87.85.gf}
% insert suggested keywords - APS authors don't need to do this
%\keywords{}

%\maketitle must follow title, authors, abstract, \pacs, and \keywords
\maketitle

% body of paper here - Use proper section commands
% References should be done using the \cite, \ref, and \label commands
%\section{}
% Put \label in argument of \section for cross-referencing
%\section{\label{}}
%\subsection{}
%\subsubsection{}

Rayleigh-B\'enard (RB) flow, in which the fluid is heated from below and cooled from above, is a paradigmatic representation of thermal convection, with many features that are of interest in natural and engineering applications. Examples range from astrophysical and geophysical flows to industrial thermal flows \cite{ahl09,loh10,chi12}. When the temperature difference between the two plates (expressed in dimensionless form as the Rayleigh number Ra) is large enough, the system is expected to undergo a transition from the so-called ``classical regime" of turbulence, where the boundary layers (BLs) are of laminar type\cite{sun08,zho10,zho10b,pui16}, to the so-called ``ultimate regime'', where the BLs are of turbulent type, as first predicted by Kraichnan \cite{kra62} and later by others \cite{spi71,gro00,gro01,gro11,gro12}. In the classical regime, the Nusselt number Nu (dimensionless heat transfer) is known to effectively scale as Ra$^\beta$, with the effective scaling exponent $\beta \leq1/3$ \cite{gro00,gro01,ste13,mal54,pri54}. Beyond the transition to the ultimate regime, the heat transport is expected to increase substantially, reflected in an effective scaling exponent $\beta > 1/3$~\cite{kra62,ahl09,gro11}. 
%$\beta = 1/3$, predicted in the early times of RB research \cite{}, is the threshold effective scaling exponent. 

Hitherto, the evidence for the transition to the ultimate regime has only come from experimental measurements. In fact, the community is debating at what Ra the transition starts and even whether there is a transition at all. For example, Niemela \& Sreenivasan \cite{nie10} observed that $\beta$ first increases above 1/3 around $\textrm{Ra}\approx10^{14}$ and then decreases back to 1/3 again for $\textrm{Ra}\approx10^{15}$. Subsequently, Urban {\it et al.} \cite{urb14} also reported $\beta \approx 1/3$ for $\textrm{Ra}=[10^{12},10^{15}]$. However, Chavanne {\it et al.} \cite{cha97,cha01} found that the effective scaling exponent $\beta$ increases to 0.38 for $\textrm{Ra}>2\times10^{11}$. In the experiments mentioned above, low temperature helium was used as the working fluid and Prandtl number Pr changes with increasing Ra. In contrast to helium, SF$_6$ has roughly pressure independent Pr. This allows He {\it  et al.}~\cite{he12,he12a} to achieve the ultimate regime more conclusively. They observed a similar exponent 0.38, but this exponent was found only to start at a much higher Ra $\approx 10^{14}$ (the transition starts at Ra $\approx 10^{13}$). This observation is compatible with the theoretical prediction~\cite{gro00,gro01} for the onset the ultimate regime. It is also consistent with the theoretical prediction of Refs. \cite{kra62,gro11}, according to which logarithmic temperature and velocity BLs are necessary to obtain an effective scaling exponent $\beta \approx 0.38$ for that Ra.

The apparent discrepancies among various high Ra RB experiments have been attributed to many factors. The change of Pr, the non-Boussinesq effect, the use of constant temperature or constant heat flux condition, the finite conductivity of the plates, and the sidewall effect can all play different roles \cite{ahl09,stevens2011}. Direct numerical simulations (DNS), which do not have these unavoidable artefacts as occurring in experiments, can ideally help to understand the transition to the ultimate regime, with the strict accordance to the intended theoretic RB formulations. Unfortunately, high Ra simulations in three dimensions (3D) are prohibitively expensive~\cite{shi10, ste10}. The highest Rayleigh number achieved in 3D RB simulations is $2 \times 10^{12}$~\cite{stevens2011}, which is one order of magnitude short of the expected transitional Ra. Two-dimensional (2D) RB simulations, though different from 3D ones in terms of integral quantities for small Pr~\cite{sch04,poe13}, still capture the many essential features of 3D RB \cite{poe13}. Consequently, in recent years, 2D DNS has been widely used to test theories, not only for normal RB \cite{hua13,zha17}, but also for RB in porous media \cite{hew12}. Although also expensive at high Ra, now we have the chance to push forward to $\textrm{Ra} =10^{14}$ using 2D simulations as we will show in this manuscript.  

%However, due to the complexity of the experiments, it remains indeterminate whether the mean profiles (velocity and temperature) are indeed logarithmic or not.
%Specifically, for temperature, there exists a few measurements of the logarithmic profiles locally in the near-sidewall regions of RB cells \cite{ahl12,ahl14}.  In contrast, for velocity, there is almost no evidence for the existence of logarithmic BLs owing to certain experimental challenges.
%Velocity measurements are difficult to obtain for the following reasons. 
%For instance,  in cylindrical cells with aspect ratio $\Gamma  = \mathcal{O}(1)$, the mean velocity profile cannot be quantified easily due to the absence of a stable mean roll structure [24].
%The absence of a stable mean roll structure in RB cells with aspect ratio $\Gamma \sim \mathcal{O}(1)$ hinders the quantification of the mean velocity profiles~\cite{stevens2011}. 
%In situations where stable mean rolls do exist (e.g. narrow rectangular cells), the highest Ra available are still far below the critical Ra at which logarithmic velocity BLs can manifest themselves \cite{sun08,pui16}.

Another advantage of DNS as compared to experiment is that velocity and temperature profiles can be easily measured, to check whether they are logarithmic in the ultimate regime, as expected from theory. Specifically, for the temperature, only a few {\it local} experimental measurements were available in the near-sidewall regions of RB cells, which showed logarithmic profiles \cite{ahl12,ahl14}. Even worse, for velocity, there is almost no evidence for the existence of a logarithmic BL, due to the experimental challenges.
%Velocity measurements are difficult to obtain for the following reasons. 
For instance,  in cylindrical cells with aspect ratio $\Gamma  = \mathcal{O}(1)$, the mean velocity profile cannot be easily quantified because of the absence of a stable mean roll structure [24].
%The absence of a stable mean roll structure in RB cells with aspect ratio $\Gamma \sim \mathcal{O}(1)$ hinders the quantification of the mean velocity profiles~\cite{stevens2011}. 
In situations where stable rolls do exist (e.g. narrow rectangular cells), the highest Ra available are still far below the critical Ra at which logarithmic velocity BLs can manifest themselves \cite{sun08,pui16}.

As numerical simulations provide us with every detail of the flow field which might be unavailable in experiments, they also enable us to reveal the links between the global heat transport and the local flow structures.
%Numerical simulations are a powerful tool to quantify the global and local flow features simultaneously.
%Numerical simulations can enable us to link the global heat transport with the local flow features. 
A few attempts (both 2D and 3D) have been made in the classical regime, in which logarithmic temperature BLs were detected, by selectively sampling the regions where the plumes are ejected to the bulk \cite{ahl12,poe15prl}.
% and similar to the experimental observations of Refs.~\cite{ahl14,wei14}.
However, it is still unclear how these local logarithmic BLs contribute to the attainment of the global heat transport enhancement during the transition to the ultimate regime. 

In this work, for the first time in DNS we do observe a transition to the ultimate regime in 2D, namely at $\textrm{Ra}^*=10^{13}$, similar as in the 3D RB experiments of Ref. \cite{he12}. DNS also provides first evidence that the mean velocity profiles follow the log-law of the wall, in analogy to other paradigmatic turbulent flows, e.g. pipe, channel, and boundary flows \cite{per82,mar10b,smi11}. Further, we explore the link between the local and global quantities to reveal the mechanism leading to the increased scaling exponent beyond the transition.
%Further, we observe that the spatial extent of the plume ejecting regions do not grow despite the increase in Ra. In fact, the effective scaling exponent is enhanced due to the increase in the contribution from the plume ejecting regions.

\begin{figure}[!bp]
	
	\includegraphics[width=.48\textwidth]{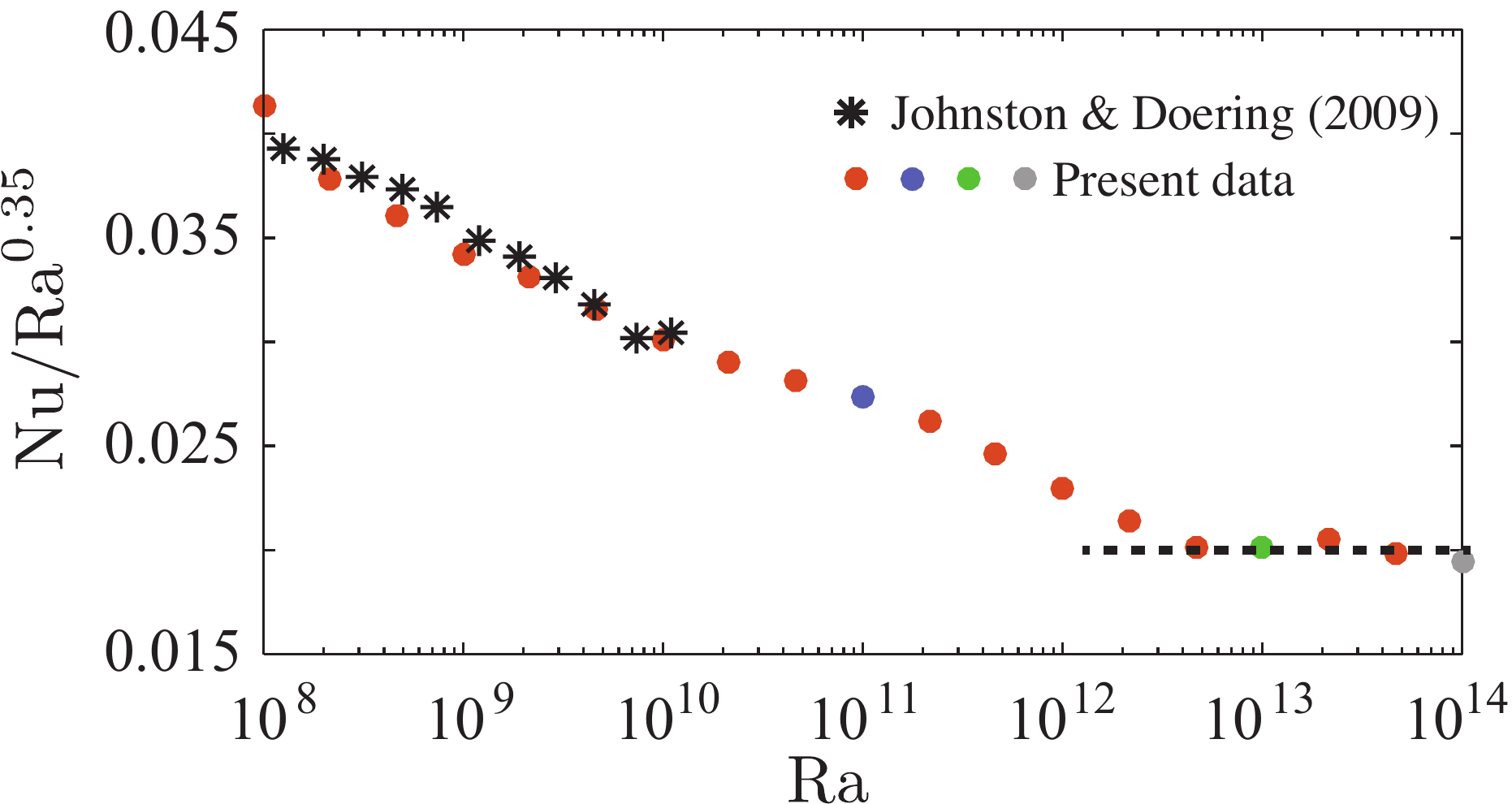}
	
	\caption{Nu(Ra) plot compensated by Ra$^{0.35}$. A clear transition can be seen at Ra = 10$^{13}$, as evident from the plateau. The data agree well with the previous results in the low Ra regime \cite{joh09}. The flow structures of the three colored data points (blue for $\textrm{Ra}=10^{11}$, green for $\textrm{Ra}=10^{13}$, grey for $\textrm{Ra}=10^{14}$) are displayed in Fig. \ref{fig1}.}
	\label{fig2}
\end{figure}

\begin{figure}
	
	\includegraphics[width=0.48\textwidth]{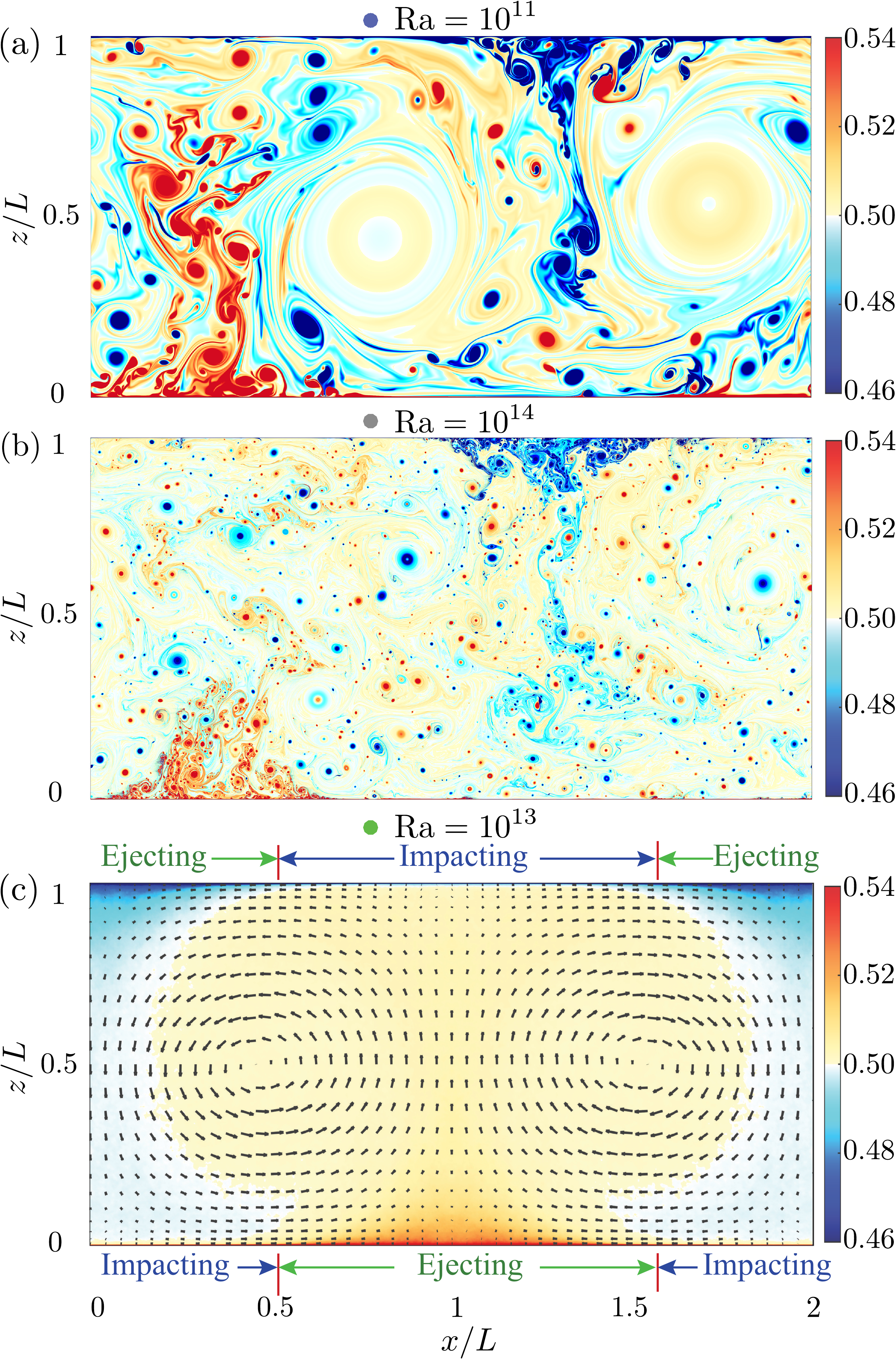}
	
	\caption{The instantaneous temperature fields for (a) $\textrm{Ra}=10^{11}$ and (b) $\textrm{Ra}=10^{14}$. The corresponding movies are shown in Ref. \cite{suppl}. (c) The Mean temperature and velocity field for $\textrm{Ra}=10^{13}$. The contours represent the mean temperature field, while the vectors show the direction of the velocity, scaled by its magnitude. The plate surfaces have been divided into equal-sized plume ejecting and impacting regions. The three plots share the same color map.}
	\label{fig1}
\end{figure}

The simulations have been carried out using a well validated second-order finite-difference code \cite{ver96,poe15cf}. %No-slip conditions were used for the velocity, temperature is constant for bottom and top plates, and horizontal sidewalls are periodic. 
The two control parameters are $\textrm{Ra}=\alpha g \Delta L^3/(\nu \kappa)$ and $\textrm{Pr}=\nu/\kappa$, with $\alpha$ being the thermal expansion coefficient, $g$ the gravitational acceleration, $\Delta$ the temperature difference across a fluid layer of depth $L$, $\nu$ the kinematic viscosity, and $\kappa$ the thermal diffusivity. In the simulations, Pr were fixed at 1 and aspect ratio $\Gamma \equiv W/L$ were fixed at 2, where $W$ is the width of the domain. With this $\Gamma$, it has been found that the heat flux approximates the heat flux at infinite aspect ratio \cite{joh09}. The boundary conditions were non-slip for velocity, constant temperature for the bottom and top plates, and periodic horizontally. 
%$\Delta$  and the free fall velocity $U=\sqrt{\alpha g \Delta L}$ are used to nondimensionalize other flow quantities. 
$\textrm{Nu}$ was calculated from the relation $\textrm{Nu}=\sqrt{\mathrm{Ra}\mathrm{Pr}}\left <u_z \theta \right >_{A,t} -\left <\partial_z \theta \right >_{A,t}$, with $u_z$ being the vertical velocity, $\theta$ the temperature, and $\left <\cdot \cdot \cdot \right>_{A,t}$ the average over a horizontal plane and time. All the cases were well resolved. At the highest  $\textrm{Ra}= 10^{14}$, we used a grid with $20480\times10240$ mesh points. For details of the simulations, we refer to Ref. \cite{suppl}.

%Although the spatial extent of plume ejecting region does not grow with increasing Ra, when the ejecting regions dominate over the impacting regions in contribution to heat transfer, we observe the transition.  

%Although the simulations were two dimensional, they still entail several of the key ingredients of the transition phenomenon. For instance, the effective scaling exponents in the classical regime for 2D and 3D have been found to be identical~\cite{several}. Also, two dimensional simulations have revealed the existence of logarithmic BLs~\cite{several...}.

%\begin{figure} [!b]
	
%	\includegraphics[width=.48\textwidth]{3.eps}
	
%	\caption{Mean temperature and velocity field for $\textrm{Ra}=10^{13}$. The contours show the mean temperature and the vectors the direction of the velocity, scaled by its magnitude. The plate surfaces have been divided into plume ejecting and impacting regions.}
%	\label{fig3}
	%	\vspace{-.75 cm}
%\end{figure}

We begin by looking at the heat transport Nu as a function of Ra.  In Fig.~\ref{fig2}, we show Nu(Ra) compensated with Ra$^{0.35}$, for the range Ra=[$10^8, 10^{14}$]. Up to Ra = 10$^{11}$ (blue symbol), the effective scaling is essentially the same ($\beta  \approx 0.29$) as  has been already observed~\cite{joh09,poe13,poe14} in the classical regime where the BLs are laminar \cite{zho10,zho10b}. This trend continues up to the transitional Rayleigh number Ra$^* = 10^{13}$ (green symbol). Beyond this, we witness the start of the transition to the ultimate regime, with a notably larger effective scaling exponent $\beta \approx 0.35$, as evident from the plateau in the compensated plot.

Next, to appreciate how the flow structures are different before and beyond the transition (Ra$^*$), we show the respective instantaneous temperature fields (see Fig.~\ref{fig1}). The top panel presents a relatively low Ra $= 10^{11}$ (below Ra$^*$), while the middle panel shows a high Ra $= 10^{14}$ (beyond Ra$^*$). At low Ra, intense large scale rolls (LSR) are clearly visible. In comparison, at high Ra, the LSR, although still evident, contains much weaker and smaller structures. Interestingly, even at the highest Ra, the temperature field still has both plume ejecting and impacting regions. Additionally, these observations indicate that the spatial extent of plume ejecting regions do not grow in spite of the increase in Ra.

%we do not see a growth in the spatial extent of plume ejecting regions, in contrast to the perception that a gradual increase in the fraction of plume ejecting regions causes the transition. 

\begin{figure}[!htbp]
	
	\includegraphics[width=0.45\textwidth]{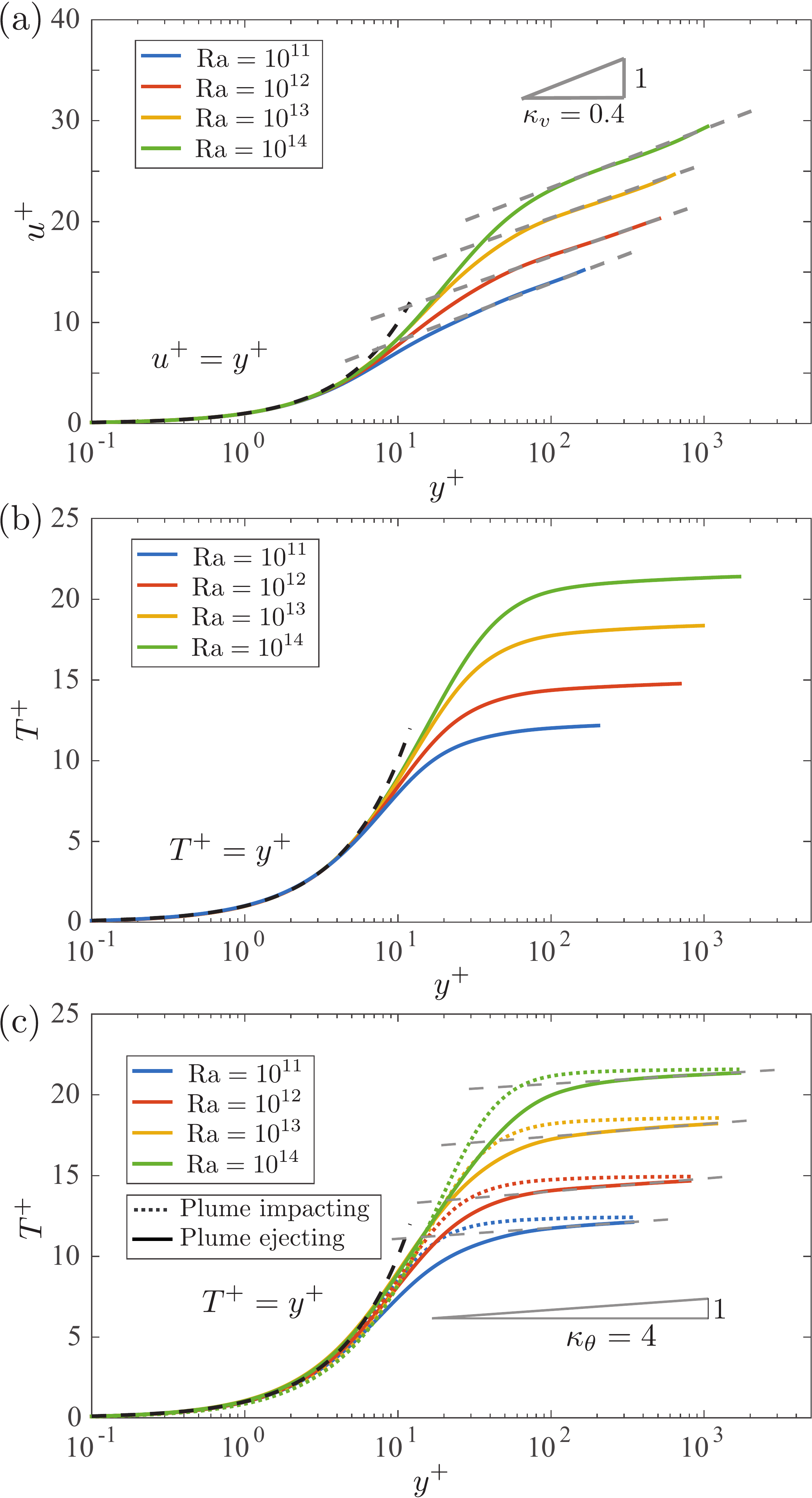}
	
	\caption{Mean velocity (a) and temperature (b) profiles in wall units ($u^+$ for velocity, $T^+$ for temperature, and $y^+$ for wall distance) at four Ra. The dashed lines show the viscous sublayer behavior and the log-layer behavior. A log-layer is seen for the velocity (with inverse slope $\kappa_v=0.4$), but not for the temperature. (c) Local temperature profiles averaged in plume ejecting and impacting regions (see Fig. \ref{fig1}(c) for definitions). The dashed lines again show the viscous sublayer behavior and the log-layer behavior. A log-layer is seen for the temperature in the plume ejecting regions (with inverse slope $\kappa_\theta=4.0$), but not in impacting regions. }
	\label{fig4}
\end{figure}

We now focus on the mean temperature and velocity fields at the transitional Ra. Remarkably, even after 500 dimensionless time units, the flow domain still shows a stable mean roll structure, i.e. the rolls are pinned with clearly demarcated plume ejecting and impacting regions (see Fig.~\ref{fig1}(c)). The mean temperature and velocity fields display horizontal symmetry, which enables us to average them over a single LSR instead of the whole domain (as the velocity averaged horizontally for the whole domain will be zero).

Figure~\ref{fig4}(a) shows the temporally and spatially averaged velocity profiles, performed  on one single LSR. We plot the profiles in dimensionless wall units, in terms of $u^+$ and $y^+$, where $u^+=\langle u \rangle_{x,t}/u_\tau$ and $y^+=zu_\tau/\nu$. Here $u_\tau$ is the friction velocity $u_\tau=\sqrt {\nu \partial_z \langle u \rangle_{x,t} |_{z=0}}$~\cite{pop00}. Similar to channel, pipe, and boundary layer flows, we can identify two distinct layers: a viscous sub-layer where $u^+ = y^+$, followed by a logarithmic region, where the velocity profile follows $u^+ = \frac{1}{\kappa_v} \ln{y^+} + B_v$ \cite{pop00}. The inverse slope gives $\kappa_v=0.4$, which is remarkably close to the K\'arm\'an constant in various 3D canonical wall-bounded turbulent flows \cite{mar10b,smi11}. However, the parameter $B_v$ varies with Ra. With increasing Ra, the logarithmic range grows in spatial extent, until at Ra$^*=10^{13}$, it spans one decade in $y^+$. Similarly, we express the temperature profile in wall units $T^+=(T_b-\langle T \rangle_{x,t})/T_\tau$, where $T_b$ is the bottom plate temperature, and $T_\tau=-\kappa \partial_z \langle T \rangle_{x,t} |_{z=0}/ u_\tau$ a characteristic temperature scale analogous to $u_\tau$ for the velocity \cite{yag79}. The mean temperature profile shows a similar viscous sub-layer $T^+ = y^+$, followed by a rather flat region, without a clear logarithmic dependence. Since the ultimate regime is associated with logarithmic profiles, the key question remains, as to why the mean temperature profile is not logarithmic despite the global scaling relations suggesting a transition in Fig.~\ref{fig2}.

To find out, we look back more closely into the flow field of Fig.~\ref{fig1}(c) where the mean flow was separated into (a) a plume ejection region, and (b) a plume impacting region. As noted earlier, the spatial extent of these regions does not grow with increasing Ra, and the mean flow field is horizontally symmetric. Therefore, the domain can be divided into plume ejection and impacting regions, enabling us to perform a conditional analysis for the temperature profiles specific to the respective regions. In Fig.~\ref{fig4}(c), we plot these profiles separately, for different Ra. All the profiles collapse into a single curve in the viscous sub-layer. Beyond the viscous sub-layer, the impacting and ejecting regions show very different behavior. For the impacting regions, the temperature profile is flat (dotted curves), and remains so for all Ra. However, for the plume ejecting regions, we observe a clear log-layer (solid curves) with a profile $T^+ = \frac{1}{\kappa_\theta} \ln{y^+} + B_\theta$, where $\kappa_\theta=4$ is the equivalent K\'arm\'an constant for temperature, and $B_\theta$ varies with Ra. Similar to the velocity profiles, the extent of the log-layer increases with Ra. At the transitional Ra$^* = 10^{13}$, it spans one decade in $y^+$. 

%This suggests that the transition point is closely linked to the plume ejection regions. 

Temperature profiles that are locally logarithmic (in plume ejecting regions) have been observed before for both the classical and the ultimate regimes \cite{ahl12,ahl14,poe15prl}. Based on this, one hypothesis regarding how the system undergoes the transition to the ultimate regime is that the fraction of plume emitting regions (or hot spots) will gradually grow with increasing Ra~\cite{poe15prl}. As speculated, the trend would continue until the entire BL becomes a hot spot, thus leading to a mean logarithmic temperature profile. Our findings indicate that here this is not the case, as even at $\textrm{Ra}=10^{14}$ plume impacting regions do not show a logarithmic temperature profile. The presence of these impacting regions makes the mean temperature profile also non-logarithmic (see Fig.~\ref{fig4}(b)).

%\begin{figure}[!tbp]
	
%	\includegraphics[width=.5\textwidth]{5.eps}
	
%	\caption{Local temperature profiles averaged in plume ejecting and impacting regions (Fig. \ref{fig1}(c)). The dashed lines show the viscous sub layer and log layer. Log law is seen for the temperature in plume ejecting regions (with slope $1/\kappa_\theta=0.25 $) but not in impacting regions. }
%	\label{fig5}
%\end{figure}

\begin{figure}[!bp]
	
	\includegraphics[width=0.47\textwidth]{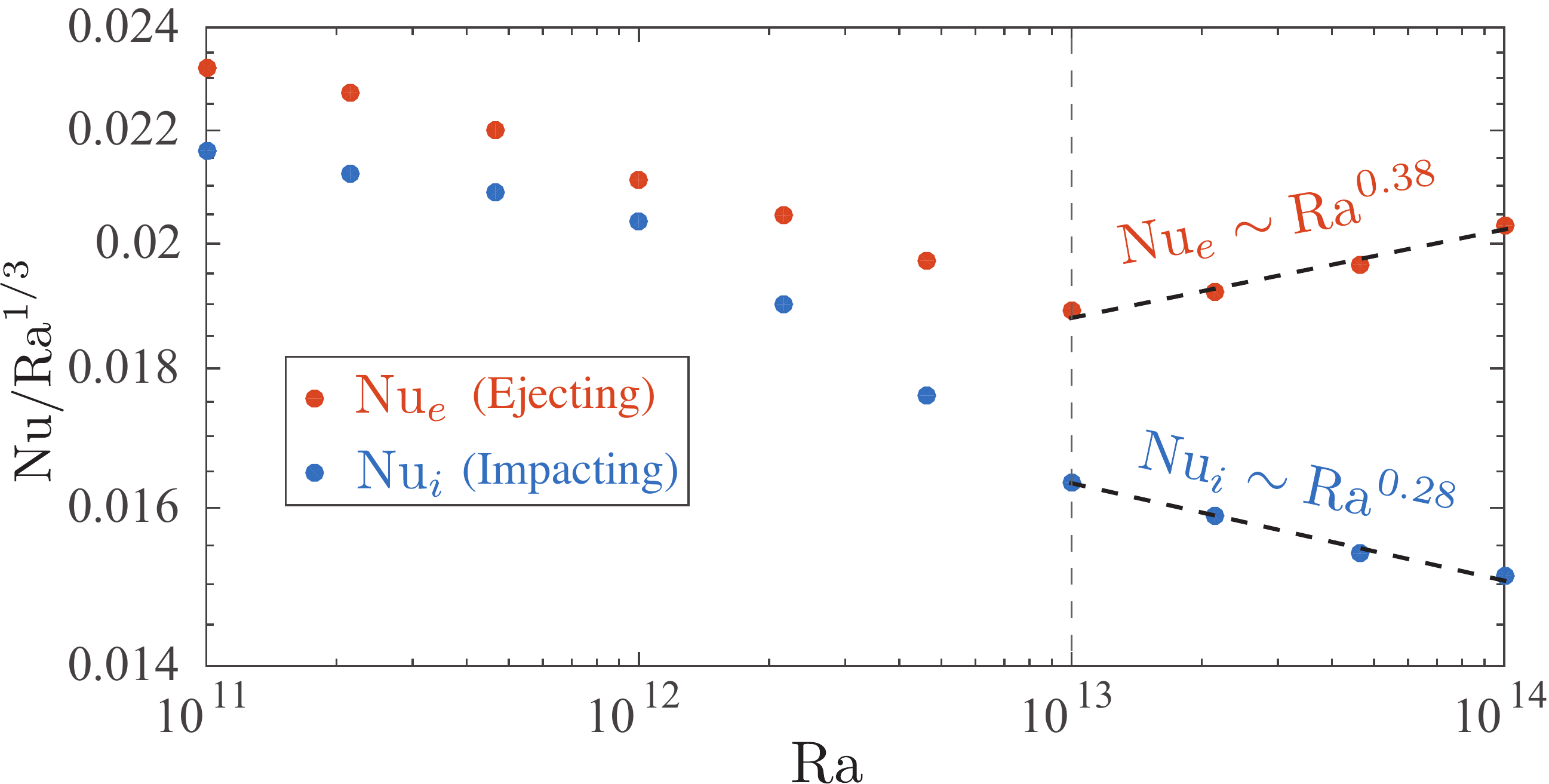}
	
	\caption{Local wall-heat-flux as a function of Ra, separately for the plume ejecting region (Nu$_e$) and the plume impacting region (Nu$_i$). At Ra$^*=10^{13}$, Nu$_e$ starts to undergo a transition to the ultimate regime with an effective scaling exponent of 0.38, while Nu$_i$(Ra) has a much smaller effective scaling exponent of 0.28. The competition between the two parts finally determines the effective global scaling exponent.}
	\label{fig6}
\end{figure}

We now explain how the global heat transport scaling can still undergo a transition to the ultimate regime, though only the local temperature profile is logarithmic, not the globally averaged one. We recall that by definition on the plate surface, Nu $ = -\left <\partial_z \theta \right >_A$. Following the observations from Fig.~\ref{fig1}(c), we compute the local Nu on the plate surface from ejecting (Nu$_e$) and impacting (Nu$_i$) regions separately. These are shown in Fig.~\ref{fig6}, compensated by Ra$^{1/3}$. Up to Ra$^*$, both Nu$_i$ and Nu$_e$ follow a similar trend, with their respective local scaling exponents $\beta_i$ and $\beta_e < 1/3$. However, beyond  Ra$^*$,  Nu$_i$ and Nu$_e$ diverge. The ejecting regions show an increased heat transport, with $\beta_e  = 0.38$, which is precisely the ultimate scaling exponent predicted for $\text{Ra} \sim \mathcal{O}(10^{14})$ with logarithmic BLs. In contrast, the impacting regions have a much lower scaling exponent $\beta_i = 0.28$. This means that the flow is partially in the ultimate regime and partially still in the classical regime. Based on these, we express the global Nusselt number, Nu = Nu$_i$ + Nu$_e$, in analogy to the Grossmann-Lohse approach \cite{gro00,gro11} wherein the dissipation rate was separated into bulk and BL contributions.  We write $\textrm{Nu} = C_i\ \textrm{Ra}^{\beta_i } +C_e \ \textrm{Ra}^{\beta_e}$, where $\beta_e$ is expected to become even larger with increasing Ra \cite{gro11}. The above expression asymptotically approaches the ultimate regime scaling when the plume ejecting regions become more and more dominant in transporting the heat with increasing Ra. Thus, with only the local temperature profile being logarithmic (in plume ejecting regions), the system can still undergo a gradual transition to the ultimate regime.

%\begin{align}
% \textrm{Nu} =  \textrm{Nu}_i + \textrm{Nu}_e,
% 
% =  A \textrm{Ra}^0.28 + B \textrm{Ra}^{0.38}
% 
%\end{align}
%\begin{eqnarray}
%
%\label{EqtofMotion}
%\end{eqnarray}

%scaling $\text{Nu} \sim \text{Ra}^{0.38}$. The spatial variation of heat flux on the bottom plate at two different Ra are given in Fig.~\ref{fig6}(b). At low Ra = $10^{11}$, the heat flux from the impacting regions dominates over the heat flux coming from the ejecting regions. As Ra increases, this behaviour is reversed, i.e most of the heat flux occurs from the plume ejecting regions. 

%\textcolor{red}{Even though 2D RB differs from three-dimensional (3D) RB
%in terms of integral quantities for finite Pr [24,32],
%the theoretical arguments for logarithmic profiles are not
%specific to 3D RB. Furthermore, the two-dimensional
%domain is more suitable to study the horizontal dependence
%of the boundary-layer profile than the three-dimensional
%domain. The locations of plume emission on the one hand
%and large-scale circulations on the other hand are more
%straightforwardly identified.}

Finally, it is worthwhile to clarify the effect of the imposed two-dimensionality on the heat transfer. As mentioned in the beginning, 2D RB is different from 3D RB. However, the effective scaling exponents observed are identical in 2D and 3D for a wide range of Ra in the classical regime \cite{poe13}, and here we found both in 2D and 3D the transition starts at Ra$^* = 10^{13}$. Furthermore, the logarithmic BLs are theoretically expected for both 2D and 3D, as the theoretical argument \cite{gro12} is built on the Prandtl equations which are 2D. Also in other 2D canonical flows, logarithmic BLs have been observed, e.g. in channel flow \cite{sam14,lvo09}. Therefore, the physical insights gained from this work are useful for understanding the transition to ultimate turbulence in both 2D and 3D flows.

In conclusion, we have used two-dimensional simulations of Rayleigh-B\'enard convection to investigate the transition to the ultimate regime of thermal convection. We followed the approach of using the local flow structures to explain the globally observed heat transfer enhancement. A transitional Rayleigh number Ra$^* = 10^{13}$ was found for the 2D RB with Pr $=1$, beyond which the mean velocity profile has a log-layer spanning one decade. However, the temperature profile is logarithmic only within the regions where plumes are ejected. The local effective Nusselt scaling exponent $\beta_e$ increases to 0.38 in the plume ejecting regions, corresponding to the ultimate regime.
%Instead, the transition occurs due to the enhancement in the local Nu scaling exponent arising from the plume ejecting regions. 
%The present study has revealed a new mechanism for the transition to the ultimate regime of thermal convection, 
The transition to the ultimate regime can be understood as the \textit{gradual} takeover of the global heat transport by the contribution from the regions of plume ejection. In future work we will extend these 2D DNS to smaller (and larger) Pr, to check the predicted Pr-dependence \cite{gro00,gro01} of the transition to the ultimate regime.

Many open questions remain, for example whether wall
roughness can trigger a transition to an asymptotic ultimate
regime, in which Nu$\sim$Ra$^{1/2}$, i.e., the logarithmic corrections
vanish. A previous study that reached Ra = $10^{12}$ has
shown that this was not yet the case \cite{zhu17b}. However, in
rough wall Taylor-Couette (TC) simulations (reaching a
Taylor number of Ta $\approx 2 \times 10^9$) and experiments (reaching
Ta $\approx 10^{12}$) we did reach the corresponding asymptotic
ultimate regime for the angular momentum transport in TC
flow thanks to the effect of pressure drag \cite{zhu18}. As the
analog to pressure drag is absent in the heat flux balance for
RB flow, such an asymptotic ultimate regime may not exist
in RB flow \cite{owe63}.

% If you have acknowledgments, this puts in the proper section head.
\begin{acknowledgments}
We thank Daniel Chung for discussions and for pointing
us to Ref. \cite{owe63}. The work was financially supported by
NWO-I, NWO-TTW, the Netherlands Center for
Multiscale Catalytic Energy Conversion (MCEC), all
sponsored by the Netherlands Organization for Scientific
Research (NWO), and the COST Action MP1305. Part of
the simulations were carried out on the Dutch national
e-infrastructure with the support of SURF Cooperative.
We also acknowledge PRACE for awarding us access to
Marconi based in Italy at CINECA under PRACE Project
No. 2016143351 and the DECI resource Archer based in
the United Kingdom at Edinburgh with support from the
PRACE aisbl under Project No. 13DECI0246.
\end{acknowledgments}

% Create the reference section using BibTeX:
%merlin.mbs apsrev4-1.bst 2010-07-25 4.21a (PWD, AO, DPC) hacked
%Control: key (0)
%Control: author (72) initials jnrlst
%Control: editor formatted (1) identically to author
%Control: production of article title (-1) disabled
%Control: page (0) single
%Control: year (1) truncated
%Control: production of eprint (0) enabled

%\bibliography{literatur}
%merlin.mbs apsrev4-1.bst 2010-07-25 4.21a (PWD, AO, DPC) hacked
%Control: key (0)
%Control: author (72) initials jnrlst
%Control: editor formatted (1) identically to author
%Control: production of article title (-1) disabled
%Control: page (0) single
%Control: year (1) truncated
%Control: production of eprint (0) enabled
%merlin.mbs apsrev4-1.bst 2010-07-25 4.21a (PWD, AO, DPC) hacked
%Control: key (0)
%Control: author (72) initials jnrlst
%Control: editor formatted (1) identically to author
%Control: production of article title (-1) disabled
%Control: page (0) single
%Control: year (1) truncated
%Control: production of eprint (0) enabled
%

%merlin.mbs apsrev4-1.bst 2010-07-25 4.21a (PWD, AO, DPC) hacked
%Control: key (0)
%Control: author (72) initials jnrlst
%Control: editor formatted (1) identically to author
%Control: production of article title (-1) disabled
%Control: page (0) single
%Control: year (1) truncated
%Control: production of eprint (0) enabled

\end{document}